# Putting Mind Back into Nature: A Tribute to Henry P. Stapp

David E. Presti
University of California, Berkeley
Department of Molecular and Cell Biology
Berkeley, CA 94720-3200 USA
Email: presti@berkeley.edu

**A contribution to a collection of essays honoring Henry P. Stapp
on the occasion of his 90th birthday in Spring 2018.**

**Abstract:** Henry P. Stapp has for 60 years been a leader – perhaps *the* leader – in exploring the role of mind/psyche/consciousness/experience in the ontology of quantum mechanics. Henry's contention is that the very structure of quantum mechanics *implies* a central and irreducible role for mind: an *experiential* aspect of nature distinct from that of the physical matter and energy described by the dynamical equations of physics. The task then becomes to generate trans-disciplinary interest in exploring this thesis, and in particular to seek connections with neuroscience and with empirical psychology.

In 1959, Henry Stapp wrote an essay titled "Mind, Matter and Quantum Mechanics," framing a project that would occupy him for decades to come. The concluding paragraph states that:

> "Quantum theory is based on concepts radically different from the classical ones and it is a natural synthesis of various antithetical viewpoints that arise from the classical way of thinking. Primarily it is a synthesis of the idealistic and materialistic world views. To some extent it also reconciles the monistic and pluralistic attitudes, provides a natural understanding of creation, and permits a reconciliation of the deterministic aspects of nature with the action of will."

Although the essay was never published, Henry applied the title more than 20 years later to a paper in *Foundations of Physics* (1) and then again to his first book (2).

MIND, MATTER AND QUANTUM MECHANICS

by H. P. STAPP

Zurich, 1959

Figure 1: Title page from Henry Stapp's 1959 manuscript.

While several of the early quantum physicists were interested in the possible role of mind in the ontology of quantum mechanics, it has been Henry Stapp who more than anyone has explored and developed this territory. Starting from von Neumann's description of the measurement problem (3), Henry arrived at the conclusion that mind/consciousness – an *experiential* aspect of nature distinct from that of the physical matter and energy described by the dynamical equations of physics – is necessarily implied by the very structure of quantum mechanics. That is, according to Henry: consciousness is central to a coherent interpretation of the measurement problem in quantum mechanics; it cannot be ignored; most interpretations of the measurement problem (e.g., many worlds), while they may be interesting, are evading an issue that ultimately must be confronted (4).

For good reasons, physicists may be reluctant to venture into explorations of the nature of mind/consciousness, believing it, reasonably, to be beyond the boundaries of their discipline and of their expertise. But Henry boldly forged ahead, hoping that biologists, neuroscientists, and philosophers would also engage and contribute their expertise to moving the discussion forward.

Thus, working in the foundations of physics, Henry honored the role of mind in nature, indeed he placed mind solidly in a central and necessary role: you simply can't have physics without it.

*Backstory, briefly*

Henry Pierce Stapp was born in Ohio, and at some point during his high-school years, was particularly inspired by a book titled *An Experiment With Time*, first published in 1927, the year before Henry's birth. The book was written by J. W. Dunne, a British military professional, aeronautical engineer, and philosopher, and was concerned with the nature of time and the possibility of precognition – the experience of something that has not yet taken place: seeing or feeling the future. The book made a big impact on Henry.

After graduating from the University of Michigan, and completing a doctorate at the University of California, Berkeley, Henry went in 1958 to Zurich, Switzerland to work with the legendary Wolfgang Pauli on quantum physics. In addition to his ongoing contributions to conventional quantum physics, Pauli, notably, had spent nearly three decades plumbing the depths of the human psyche and connections between mind and world, together with his psychiatrist-become-colleague-and-coauthor, Carl Gustav Jung (5, 6).

Pauli was certainly someone with whom Henry might have enjoyed a productive exploration of the undiscovered territory where mind and world enfold. But, alas, it was not to be, for shortly after Henry arrived in Zurich, Pauli died from complications of a rapidly growing pancreatic tumor. He was only 58 years old.

Henry wrote up his thoughts on the subject of mind, matter, and quantum physics at the time – the aforementioned unpublished manuscript – and returned to Berkeley and a position as a mathematical physicist at Lawrence Berkeley Lab. In the decades that followed, Henry continued to think deeply about the connections between mind, matter, and fundamental physics, writing dozens of papers on the subject: the measurement problem in quantum mechanics, the implications of Bell's theorem, consciousness viewed in light of quantum physics, even papers on free will and on precognition.

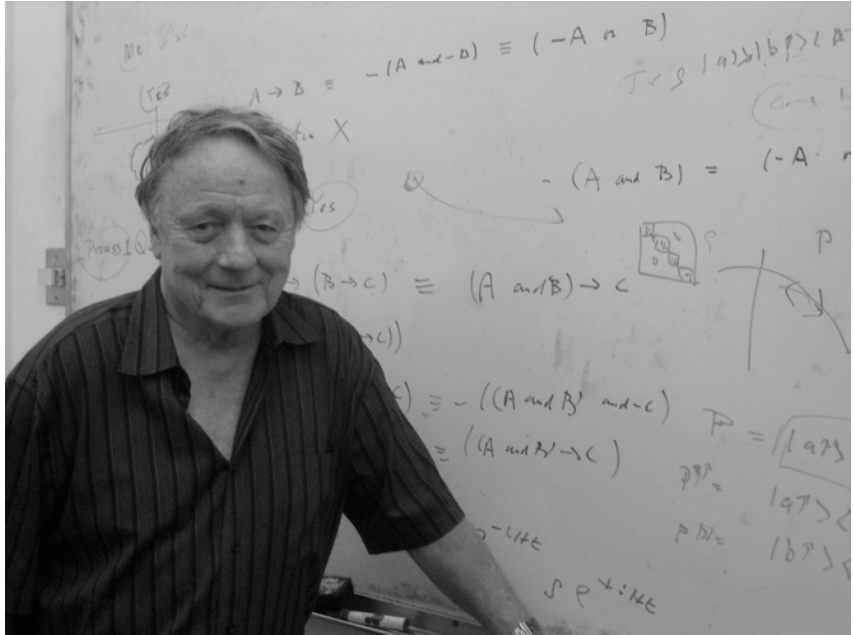

Figure 2. Henry Stapp in his office at the Lawrence Berkeley National Laboratory. 2009. Photo by David Presti.

In 1972 Henry published in the *American Journal of Physics* a beautiful paper aimed at clarifying the ideas articulated by Niels Bohr and others collectively referred to as "The Copenhagen Interpretation" (7). I read this paper when I was an undergraduate student studying physics at Butler University in Indiana. Thirty years later, when I was on the neurobiology faculty at UC Berkeley, I met Henry for the first time, at a small dinner party at his home in Berkeley, orchestrated by he and his wife Olivia. Several years after that we connected again, when we were both fortunate to be members of a trans-disciplinary academic research group at the Esalen Institute – a group dedicated to expanding the frontiers of consciousness science into radically new territory. The Esalen group produced two major books: *Irreducible Mind: Toward a Psychology for the 21$^{st}$ Century* (8) and *Beyond Physicalism: Toward Reconciliation of Science and Spirituality* (9). Henry contributed indirectly to the first book, and authored a chapter summarizing his perspective in the second (10).

*Moving forward in a science of consciousness*

All we know comes to us via our experience, and from the regularities observed through this conscious experience – this awareness – we extrapolate the existence of an objective material world and its properties. It is remarkable how well this inferred reality is self-consistent and works – how much we have been able to achieve by objectifying nature. As Einstein so aptly put it: "the eternally incomprehensible thing about the world is its comprehensibility" ["*Das ewig Unbegreifliche an der Welt ist ihre Begreiflichkeit.*"] (11).

Within this objectified world, we consider what is real to be matter and energy, particles and fields – things described by the dynamical equations of physics. And within this framework, living organisms are understood as structures composed of atoms and molecules, organized so as to utilize energy to maintain their stability. The diversity of life is considered to have developed over billions of years via biological evolution, and humans and our capacities are understood as part of this grand scenario. Within this framework mind/consciousness is considered something that emerges from the physical properties of complex nervous systems, such as are found in humans. Sentience is ultimately assumed to be explainable (although we presently have no idea how) in terms of the underlying properties of matter and energy.

In his paper on the Copenhagen interpretation – invoking the great American scientist of the mind William James – Henry stated that "if we want to know what it means for an idea to agree with a reality we must first accept that this reality lies in the realm of experience." That is, Henry begins with the premise that what we refer to as reality comes via our experience, our consciousness, our mind.

Following Descartes, pre-quantum (classical) physics removed any discussion of mind/consciousness/experience from the program of physical science. The physics that developed from this view gave rise to an explanatory framework of enormous power and utility. However, at the beginning of the 20$^{th}$ century, this framework was found incapable of accounting for the observed properties of atoms, molecules, and light. Quantum mechanics was created in order to accommodate these properties.

To Henry, a coherent interpretation of quantum mechanics requires that experience (mind/consciousness) be as primary as the physical system described by the dynamical equations (e.g., the Schrödinger or Dirac equations). Henry developed this idea in multiple publications over the decades, and his hope was that traction within a trans-disciplinary community of scholars would eventually develop.

*Connecting quantum mechanical processes with neural correlates of consciousness*

Mind/consciousness is not a phenomenon that can be tackled solely by physics. Working within our current scientific framework, the bodies of knowledge from biology, neuroscience, and psychology – at least – will be important. Can ways be imagined that connect neural processes related to mental/experiential events/qualities (NCCs: neural correlates of consciousness) with quantum processes in the brain and body?

This was an area I explored with Henry and our late colleague neuroscientist Walter J. Freeman III (1927-2016) at UC Berkeley. Freeman had for fifty years been at the forefront of an area of investigation he helped to create: cortical neurodynamics. Neurodynamics endeavors to understand the operation of the cerebral cortex not in terms of relatively limited numbers of interconnected neurons and circuits, but rather as the collective activity of very large numbers of neurons – millions, hundreds of millions, billions.

The cerebral cortex is a uniquely complex structure. Neurons and glia, together with all their multitudinous axonal and dendritic fibers and astrocytic processes, are densely packed together, connected by enormous numbers of electrical and chemical synapses, and driving local and large-scale activity also via the interactions of electromagnetic fields (ephaptic coupling) generated by the flowing charges associated with nerve signaling (12). This unique structure – the cortical neuropil – is continuously active at a very high level (13), generating global electromagnetic oscillatory activity that can be measured with electroencephalography (EEG) (14).

Freeman proposed that the cerebrocortical neuropil be described as a unified system capable of undergoing phase transitions into states of global coherence, where aspects of neural activity – at least in part measurable with EEG – are brought together in synchrony across large regions of the cerebral cortex. The nature of these cooperative states depends upon all the complexity of existing connections in the cortex, a vast network that has been assembled over a lifetime of experience. In this way, the cooperative synchrony of the neuropil functions to access memories related to present states of activity – and these memories inform the evolving experience of the perceptions that develop in association with sensory stimuli (15, 16).

Furthermore, Freeman and his physicist colleague Giuseppe Vitiello proposed that these cortical phase transitions can be described as something akin to Bose-Einstein condensates in quantum physics (17). The complex biological and electrodynamic properties of cortical neuropil suggest that it may be viewed as a new, uniquely biological, state of matter – one that allows quantum physical phase-transitions to have macroscopic impact, similar to what happens in a superfluid. Moments of consciousness – experienced as a perception, or thought, or feeling – are hypothesized to be associated with these condensation events in the cortical neuropil.

Might some connection be made between these phase transitions in cortical EEG signals and the mind-mediated quantum collapse process discussed by Henry? Might this provide a neurobiological connection that would expand the scientific conversation on the mind-matter relation? For several years I coordinated periodic meetings with Henry and Walter Freeman, exploring possible connections between their respective approaches to the mind-matter relationship.

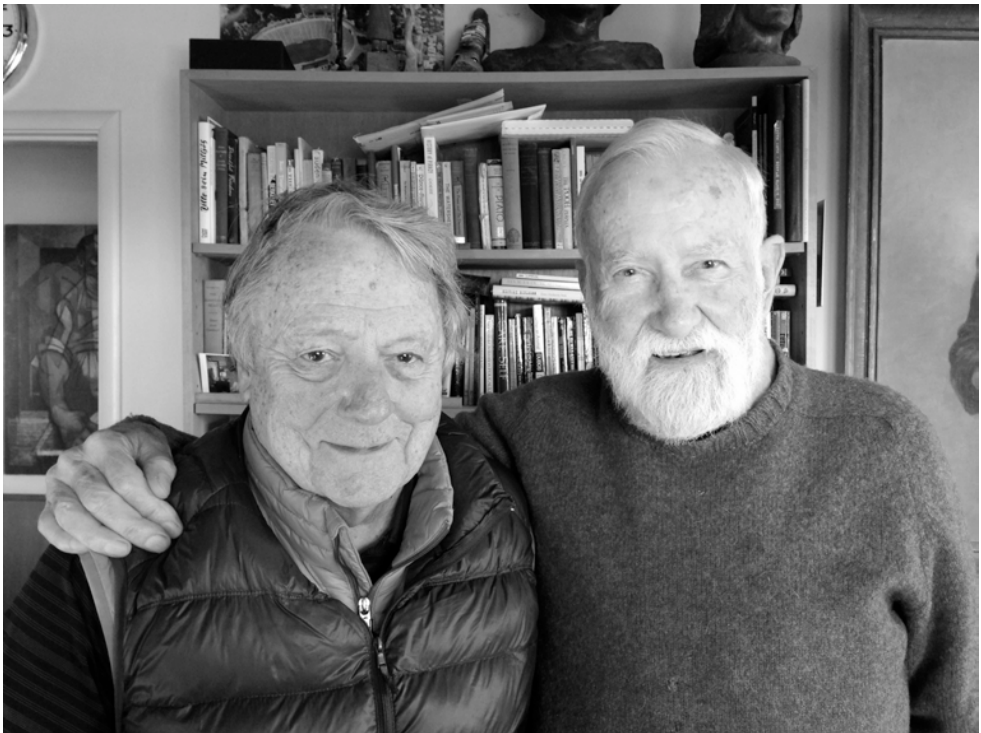

Figure 3. Henry Stapp and Walter Freeman in Berkeley.
2015. Photo by David Presti.

*A radically empirical approach to mind*

In thinking about how to expand a science of consciousness into new territory, I begin with one important premise: that there is an inescapable bidirectional relational enfolding of mind and world. We know the world only through our conscious mind. From these perceptions, thoughts, and feelings we construct explanatory frameworks that organize our experience. And from this we have developed our physical and biological science. Then, via our science, we conclude that the world manifests the conditions in which our bodies and brains and mind/consciousness occur. One might say that mind and world, mind and matter, interdependently co-create one another.

     Given this, how might we continue to expand a science of mind? A 3-part empirical framework articulated 130 years ago by William James provides a useful guide: investigate behavior (psychology), biological underpinnings or correlates (neuroscience), and mental experience (introspection or phenomenology) (18). This program is further informed by a radically inclusive empiricism: honoring all the data, even what is weird and inexplicable. Furthermore, even the metaphysical framework is questioned and considered revisable.

     Very relevant to this is investigation of phenomena falling under the broad category of psi or paranormal – phenomena involving information transfer between mind and world despite no known mechanism by which this could take place. Examples include telepathy, clairvoyance, psychokinesis, and precognition. It is not widely appreciated that laboratory studies of these phenomena have been conducted for over a century (19, 20). Also investigated are spontaneous paranormal experiences, the most robust of which are often reported in the context of powerful, emotionally-impactful circumstances: for example, near-death experiences, apparitions associated with death, and small children who have spontaneously related detailed information about another life (21). It could be that powerful emotionality disturbs in some way the fabric of ordinary reality, allowing greater access to the kinds of information transfer termed "paranormal." [It is worth noting that the term "paranormal" was introduced in the early 20th century to describe empirical phenomena that were not explicable by the normal rules of experience. It was considered a legitimate topic of academic scientific research, and only later acquired the negative connotations and emotional sociocultural baggage that the term often evokes today in scientific circles.]

     While some have claimed that the existence of such anomalous phenomena is impossible according to the laws of physics, Henry takes a broader, more open-minded and humble approach: "That one's mind can influence one's future actions is a center-piece of orthodox QM, and much of the paranormal could be explained if this power of mind were to extend beyond the perceiving subject's body" (22).

     That mind may in some way extend beyond the brain and body – connecting personal conscious awareness with more extended sources of information in the world – is a powerful and compelling hypothesis that could account for many currently inexplicable phenomena. It is also a hypothesis that can be subjected to rigorous scientific investigation, both empirical and theoretical.

*Coda*

Science evolves by way of accumulation of observations, knowledge, theorizing, and so forth – within the constraints of an accepted framework, or paradigm, on how to view things.  And scientific evolution is punctuated by paradigm shifts, or revolutions: the heliocentric cosmos, the relativity of space and time, quantum physics, biological evolution by variation and selection.  Before and after such revolutions, the world is viewed very differently (23).

I believe, as did William James, and I think also Henry Stapp, that the next great paradigm shift or scientific revolution will be in cognitive science, broadly defined – that is, the study of mind in all its multifarious aspects.  I further believe we are poised on the threshold of this great punctuation, and may already be witnessing its beginnings.  When it really gets off the ground, it will be a doozy of a revolution, impacting every science from physics to biology to psychology, and impacting society and culture at large.  And potentially, hopefully, impacting the relationship between our scientific and religious traditions and institutions.  After all, what is addressed here impacts everything about who we believe we are as conscious beings and how we relate to the rest of what we consider to be reality.  I hold a very optimistic view regarding what is possible.

Perhaps the greatest lesson of biophysical science is the deep interconnectivity of all physical phenomena.  Nothing exists except in interdependent relation with everything else.  All matter was forged in the crucibles of stars or exploding stars or in the reverberations of the Big Bang itself.  Life arose out of the complexity of interacting atoms and molecules on planets conducive to the formation of certain stable molecular configurations.  The Earth's biosphere – geology, oceans, atmosphere, and all of life from viruses to microorganisms to humans to cetaceans – is a deeply interconnected and interdependent whole.  Our own bodies harbor a symbiotic symphony consisting of vast numbers of microbes, the actions of which contribute profoundly to our health and disease. We would not be alive without them.  In a very real and physically tangible way, science is showing that the material universe is truly all-one.

Perhaps the greatest lesson of quantum physics is that mind/consciousness is also central to this deep interconnectivity – irreducibly part of the interdependent mix.  Mind has causal efficacy in the physical world: not just in experimental set-ups exploring the foundations of quantum measurement, but at every moment – within, and perhaps beyond, the body.  What we choose to *think* may matter more than we can presently imagine.

As Henry puts it:

> "[This] radical shift in the physics-based conception of man from that of an isolated mechanical automaton to that of an integral participant in a non-local holistic process that gives form and meaning to the evolving universe is a seismic event of potentially momentous proportions. . . . With our physically efficacious minds now integrated into the unfolding of uncharted and yet-to-be-plumbed potentialities of an intricately interconnected whole, the responsibility that accompanies the power to decide things on the basis of one's own thoughts, ideas, and judgments is laid upon us.  This leads naturally and correctly to a concomitant elevation in the dignity of our persons and the meaningfulness of our lives" (24).

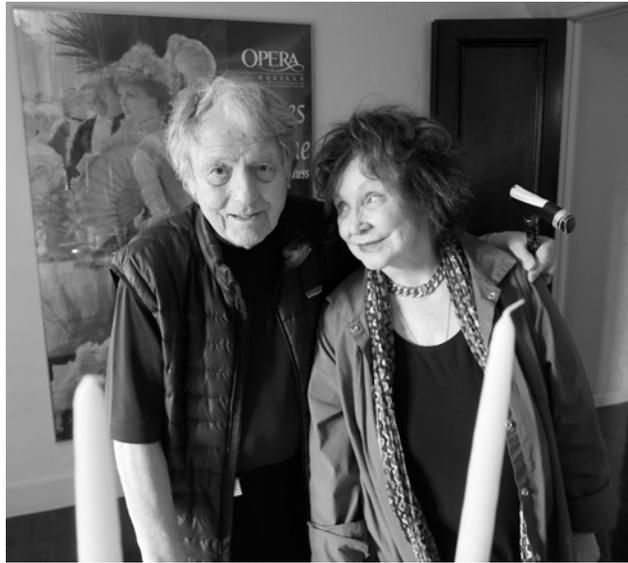

Figure 4. Henry and Olivia Stapp in Berkeley on Henry's 90th birthday: March 23, 2018. Photo by David Presti.